\documentstyle[aps,prl,twocolumn,epsf,rotate]{revtex}
\begin{document}
\flushbottom
\draft
\twocolumn[\hsize\textwidth\columnwidth\hsize\csname
@twocolumnfalse\endcsname
\title{Metallic stripe in two dimensions: stability and spin-charge separation}
\author{A. L. Chernyshev$^{1,}$\cite{perm}, A. H. Castro Neto$^{1}$, 
and A. R. Bishop$^{2}$}
\address{
$^1$Department of Physics, University of California, Riverside, CA 92521}
\address{
$^2$ Theoretical Division,
Los Alamos National Laboratory, Los Alamos, 87545 NM}
\date{\today}
\maketitle

\widetext\leftskip=1.9cm\rightskip=1.9cm\nointerlineskip\small
\begin{abstract}
\hspace*{2mm}
The problems of charge stripe formation, spin-charge separation, 
and stability of the antiphase domain wall (ADW)
associated with a stripe are addressed using an analytical approach to the
$t$-$J_z$ model. We show that a metallic stripe together with its
ADW is the ground state of the problem in the low doping regime. 
The stripe is described
as a system of spinons and magnetically confined holons 
strongly coupled to the two dimensional (2D)
spin environment with holon-spin-polaron elementary excitations filling
a one-dimensional band. 
\end{abstract}
\pacs{PACS numbers: 71.10.Fd, 71.10.Pm, 71.27.$+$a, 74.20.Mn}
]
\narrowtext
In a light of experimental evidence for incommensurate spin 
and charge order in high-T$_c$ superconductors \cite{stripesExp} the
concept of stripes, i.e. one-dimensional spatial structures of holes
condensed at an antiphase domain wall (ADW) of antiferromagnetic
(AF) spins, has been discussed extensively in the literature 
\cite{U,tJ,tJ1,stripes,arg,KE,Coulomb,WS}. In 
the physics of high-T$_c$ superconductors magnetism plays an important
role \cite{pines} and it is believed that stripe formation is due to
the strong interplay between the charge and spin degrees of freedom
\cite{arg}.
Various proposals for the driving mechanism of such a stripe phase
have been made. A large body of them emphasize the importance
of the so called {\it topological doping}, 
i.e. the breaking of the symmetry of the
ground state in order to lower the total energy of the system
when the charge carriers are introduced \cite{KE}. For the cuprates the
symmetry breaking corresponds to the change of the direction of the
staggered magnetization across the stripe, i.e. to 
the formation of an ADW.
In this context the importance of the kinetic energy of holes, 
both along the domain wall and transverse to it, has been noted in the
studies of the Hubbard and the $t$-$J$ models \cite{U,tJ,tJ1}.
The role played by the long-range Coulomb field in the ordering of the
stripes has been also discussed \cite{arg,Coulomb}. Numerical studies of the
$t$-$J$ model indicate that the (dynamic) stripe phase may be 
the ground state of the system but there is still no consensus on
this issue \cite{WS,HM}. However, the actual picture of such
topological doping is far from being complete mainly due to the lack
of microscopic insight into the problem. There is no clear
understanding of the stabilization mechanism of the stripe as well
as of the nature of low-energy excitations of the system.
Thus it is important to study a model of AF spins and mobile holes
within an approximation which is not subject to the
limitations of mean field theories or numerical studies.

In this paper we study analytically the problem of the stability of an ADW
in the anisotropic $t$-$J_z$ model. We calculate the Green's function
of a charge excitation at the ADW in a well controlled procedure.
We show that although an empty ADW (no holes) is a highly
excited state it is stabilized even at low doping concentrations 
due to the gain in hole kinetic energy. As a consequence the doped holes
become spatially confined in the direction transverse to the ADW 
and exhibit spin-charge separation features. 
 Namely, the holes 
behave like holons (charge-$1$, spin-$0$ fermions) at the ADW and
they are spin polarons (charge-$1$, spin-$1/2$ fermions) outside of
the ADW.
The confining potential in the transverse direction
leads to quantized states whose energy grows roughly $\sim J_z^{2/3}$. 
We argue that due to the 1D nature of an ADW holon even
Trugman loops \cite{Trug} are energetically costly. 
We also compare some of our results to recent DMRG calculations for the
{\it isotropic} $t$-$J$ model \cite{WS} and find very good agreement.

Our starting point is the $t$-$J_z$ model which is given by:
\begin{eqnarray}
\label{H}
{\cal H} = -t\sum_{\langle ij\rangle\sigma}(\tilde{c}^\dagger_{i\sigma}
\tilde{c}_{j\sigma}+{\rm H.c.})+ 
J\sum_{\langle ij \rangle} \bigl[S_i^z   S_j^z-
\frac{1}{4}N_iN_j \bigr]\ , 
\end{eqnarray}
where $t$ is the kinetic energy, $J$ is the AF
exchange, and $N_i=n_{i\uparrow}+n_{i\downarrow}$. 
All operators are defined in the space without
double-occupancy of the sites. The natural source of exchange
anisotropy in real systems is the spin-orbit coupling \cite{Rice}. 
Although the $t$-$J_z$ model is the
strongly anisotropic limit of the more realistic $SU(2)$ $t$-$J$ model,
it is known
from numerical and analytical studies that the low-energy physics
of holes is very similar in both models \cite{Dag}. Therefore, as long as the
hole dynamics is concerned, it should be instructive to study
the $t$-$J_z$ model which captures general properties
of the doped AFs and allows an analytical
treatment of the problem. 

According to the idea of topological doping a single charge
carrier must benefit energetically  from breaking the symmetry of the
ground state. Thus, it is natural to consider the problem of one hole
at the ADW, in order 
to obtain the low-energy variables which then can be
used for description of many-hole system. The single-hole problem in a
homogeneous AF background is very well studied
with analytical results and numerical data being in a very good 
agreement \cite{CP}. The charge quasiparticle is understood as a spin
polaron, i.e. a hole dressed by strings of spin excitations
\cite{Dag,CP}. In other words, the hole movement in a regular AF is
frustrated because of the tail of misaligned spins following the
hole. The idea that an ADW can be more favorable
configuration for holes relies on the fact that such a frustration of the
hole's kinetic energy can be avoided for movement inside the wall,
such that the hole is essentially free in the 1D structure. Then, holes
populating this 1D band can compensate the
cost of magnetic energy ($\sim J\cdot length$) of the wall. However,
this 1D energy alone ($\sim -2t$) cannot overcome the energy of the spin
polaron in the bulk 
$E_p\approx -2\sqrt{3}t$ (at $t\gg J$), especially if the energy cost for
creating an ADW is taken into account. Therefore, the 
``transverse'' hole movement away from the domain boundary must be
equally important. Moreover, it is evident that such a movement is
very similar to the hole movement in a homogeneous AF, i.e. it 
also generates strings of misplaced spins. It is easy to see
that such strings are also weaker, i.e. cost less
energy near the ADW than in the bulk. Then, the coupling of the
longitudinal and transverse hole movements for this many-body problem 
gives a true low-energy elementary excitation, 
which unifies the features of the 1D charge carrier with the
properties of the spin polaron.

\begin{figure}
\unitlength 1cm
\epsfxsize=17.cm
\begin{picture}(17,5.5)
\put(0,0.5){\epsffile{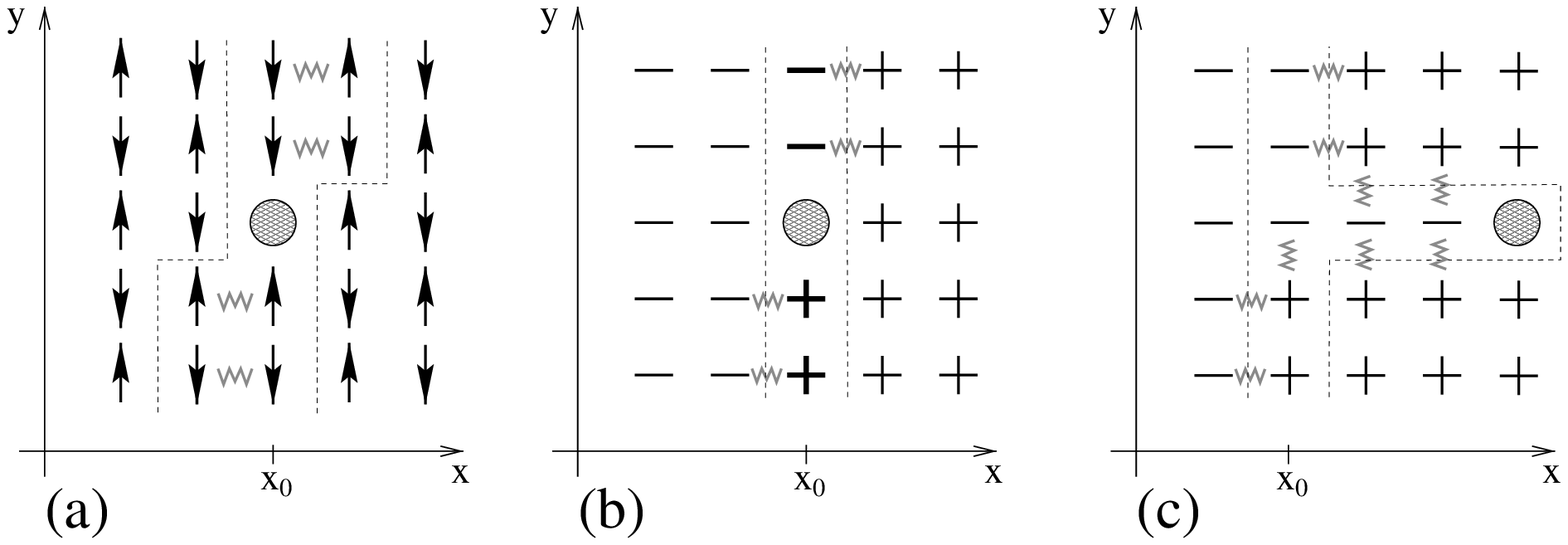}}
\end{picture}
\caption{Single hole at an ADW separating AF domains. 
Broken AF bonds are marked by strings.
(b) Same as (a). Pluses and minuses represent
staggered magnetization. (c) String of spin flips generated by the hole.}
\label{fig1}
\end{figure}

Let us consider a bond-centered domain wall with one hole injected
in it. If the hole is allowed to move along the wall it will
create a spin defect (a spinon - charge-$0$, spin-$1/2$ fermion, which
is massive in our case because of the anisotropy) and then propagate freely
as a holon. The structure shown in
Fig.\ref{fig1}a is the natural starting point for the consideration of a hole
whose movement along the stripe is free from the beginning. 
Fig.\ref{fig1}b shows the same
configuration schematically. It is clear that the domain
wall not only corresponds to an antiphase shift of the staggered
magnetization across the stripe, but also ensures that the hole moves
as a holon in the longitudinal direction.
Fig.\ref{fig1}c shows an example of a string generated by the transverse
hole movement. Notably, the first element of such a string is a spinon.
These strings lead to an effective confinement of the
hole in the direction perpendicular to the stripe and to the formation of
a spin-polaron-like cloud of spin excitations around the holon. 
This can be viewed as a generalization of 1D physics to higher
dimensions in the presence of an ADW \cite{Bob}.
 
\begin{figure}
\unitlength 1cm
\epsfxsize=7cm
\noindent
\begin{picture}(6,6.5)
\put(4,0.5){\rotate[r]{\epsffile{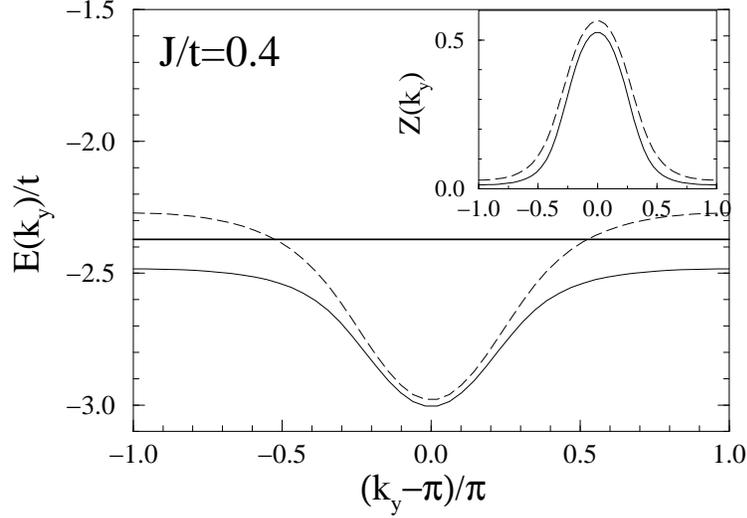}}}
\end{picture}
\caption{Energy of elementary excitation $E(k_y)$ v.s. $k_y$, 
residue of the Green's function versus $k_y$ (inset). 
Dotted lines are for path-independent string calculations, solid lines
are the results of more rigorous calculations described in the text.
Solid straight line is the energy of a spin polaron in the bulk. All
energies are relative to the energy of a static hole in a
corresponding magnetic background.} 
\label{fig3}
\end{figure}

The bare Green's function for longitudinal hole movement in Fig.\ref{fig1}b is
given by
\begin{equation}
\label{G0}
G^0_{x_0}(k_y,\omega)=[\omega-2t\cos(k_y)+i0]^{-1}\ ,
\end{equation}
so the holon band minimum is located at $k_y=\pi$, where the 
index $x_0$ corresponds
to the $x$-coordinate of the stripe. Then the renormalization of the
Green's function, as in the case of the spin polaron,
is coming from the retraceable path movements of the hole away from
ADW and back. The retraceable path
approximation is equivalent to the self-consistent Born approximation
for the self energy. Corrections due to terms
beyond this approximation are very small \cite{CP} and will
be omitted. The full Green's function is
then given by
\begin{equation}
\label{G}
G_{x_0}(k_y,\omega)=[\omega-2t\cos(k_y)-\Sigma_{x_0}(\omega)+i0]^{-1}\ ,
\end{equation}
where $\Sigma_{x_0}(\omega)$ takes the form of a continued fraction
\begin{equation}
\label{S}
\Sigma_{x_0}(\omega)=\frac{2t^2}{\omega-\omega_1-\frac{3t^2}
{\omega-\omega_1-\omega_2-\dots}}\ ,
\end{equation}
where $\omega_i$ is the energy of the $i$-th segment of the string, which
is equal to the number of broken AF bonds ($J/2$ each) associated with
the segment. In the retraceable path approximation $\Sigma_{x_0}(\omega)$ 
has no $k_y$ dependence. 
Since the energy spectrum of the elementary excitations is given
by the poles of the Green's function Eq. (\ref{G}) with the self energy
Eq. (\ref{S}), one needs to calculate $\Sigma(\omega)$ and seek 
solutions of $E(k_y)-2t\cos(k_y)-\Sigma(E(k_y))=0$. 
A standard simplification is to assume 
that the energy of the string is independent of the path of a hole and
is simply proportional to the length of the path. 
For the spin polaron this is 
plausible because only very few strings do not follow this rule. Then
an analytical solution for the self energy is given by the ratio of Bessel
functions \cite{CP}. If we assume that
$\omega_1=J/2$, $\omega_{i>1}=J$ (two broken bonds per segment of string, see
Fig. \ref{fig1}c), Eq. (\ref{S}) transforms to
\begin{equation}
\label{S1}
\Sigma(\omega)=\frac{2t^2}{\omega-J/2+\sqrt{3}t\Upsilon(\omega-J/2)}\ ,
\end{equation}
with $\Upsilon(\omega)={\cal J}_{-\omega/J}(r)/ 
{\cal J}_{-\omega/J-1}(r)$, ${\cal J}_\nu(r)$ the
Bessel function, and $r=2\sqrt{3}t/J$. 
The energy of the lowest pole of $G(k_y,\omega)$
Eq. (\ref{G}) at $J/t=0.4$ with $\Sigma(\omega)$ from (\ref{S1}) 
versus $k_y$ is plotted in Fig.\ref{fig3} (dotted line). 
The zero energy level is chosen to be equal to the energy of a static
hole in the configuration in Fig.\ref{fig1}a. This reference to the $t=0$
limit is natural to show which of the magnetic configurations is best
for optimizing the kinetic energy.

One can consider the problem more rigorously taking into account the
energy of each string exactly up to a certain length $l_c$ and
applying the path-independent assumption only for $l>l_c$. The
lowest-pole energy versus $k_y$ for such calculations with $l_c=4$ is 
shown in Fig.\ref{fig3} (solid line).
Because of the ADW, there is less energy required to create a spin flip
near it and therefore there is a subset of strings having lower
energy than the string of the same length in the bulk. 
It is worth mentioning that the energy of this
ADW elementary excitation {\it relative to the energy of the magnetic
background} is lower than the energy of the spin polaron in the bulk at
all $k_y$ (Fig.\ref{fig3}). 
For $J/t=0.4$ the gain of
the energy of the holon at the bottom of the band over the
spin polaron is about $1.5J$, that is the energy of three broken AF
bonds. 
Another informative
quantity, the residue of the Green's function $Z(k_y)$,
is shown in Fig.\ref{fig3} (inset). It gives a measure of the amount of
``bare'' holon in the wave function of the elementary excitations. 
One can see
that a significant part of the initial holon at $k_y=\pi$ resides inside
the wall and almost all its weight is transfered to strings at
$k_y<\pi/2$. Because the band is very flat at the same $k_y<\pi/2$,
the velocity of the elementary excitations is much slower than
the bare Fermi velocity $v_F^0=2t\sin(k_F)/\hbar$.
It is easy to show that the velocity at the Fermi level for our 1D band
is $v_F=v_F^0 \, Z(k_F)$. For realistic values of $J$,
$t$, and $k_F$ around $\pi/2$ one finds $\hbar v_F \sim 50-100$ meV\AA, 
which is close to the velocity of the slow mode ($\sim 35$ meV\AA)
deduced from experimental data in Ref.\cite{Bal} for the cuprates. 

\begin{figure}
\unitlength 1cm
\epsfxsize=6cm
\noindent
\begin{picture}(6.5,6)
\put(4.5,1){\rotate[r]{\epsffile{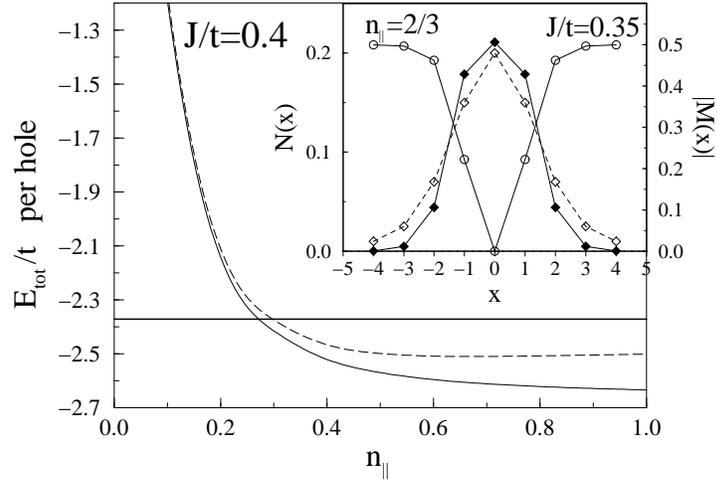}}}
\end{picture}
\caption{Total energy of the system per hole
versus $n_{\parallel}$. Horizontal line is the energy of free spin
polarons in the homogeneous AF. Dotted and solid lines are the same as
in Fig.\ref{fig3}, $J/t=0.4$. 
Inset: spatial density of holes (diamonds)
and modulus of the staggered magnetization (circles) 
across the stripe at $n_{\parallel}=2/3$, $J/t=0.35$. 
Lines are guide to the eye. 
Empty diamonds are numerical data from Ref.\protect\cite{WS}}
\label{fig4}
\end{figure}

The next step is to fill the obtained 1D band up to some Fermi 
momentum $k_F$ and study
the stability of the whole system as a function of the 1D hole
concentration $n_{\parallel}$. Obviously, the rigid band filling
neglects all effects of interaction between the carriers except 
Fermi repulsion. Such interactions would include attractive as well as
repulsive terms coming from the  nearest-neighbor hole-hole interaction,
crossing of the strings, spin-flip exchange, and kink-kink or kink-antikink
scattering of holons. However, as a first step towards the microscopic
study of the metallic stripe, the rigid-band approximations should help to 
establish the energy scale underpinning the stripe ground state.
The total energy of the stripe must include the magnetic energy paid for
the domain wall $E_w=(L_y-N_h)J/2$ and kinetic energy of holes
$E_{kin}=\sum_{k<k_F}E(k_y)$. Here, $L_y$ is the number of sites in
the $y$-direction, $N_h$ is the number of holes, $k_F=\pi n_{\parallel}$
is the Fermi momentum relative to $\pi$ (band minimum),
and $n_{\parallel}=N_h/L_y$. Thus, the total energy per hole is
\begin{equation}
\label{E}
E_{tot}/N_h=\frac{J}{2}\left(\frac{1}{n_{\parallel}}-1\right)
+\frac{1}{2\pi n_{\parallel}}\int_{\pi-k_F}^{\pi+k_F}E(k_y)dk_y \ .
\end{equation}
Results for $E_{tot}/t$ versus $n_{\parallel}$ at $J/t=0.4$ are shown in
Fig.\ref{fig4}. 
The stripe wins over the homogeneous phase when $n_{\parallel}$ is as
low as $1/4$ and then it has significantly lower energy. 
Another feature of the data is that from
$n_{\parallel}\approx 0.5$ to $n_{\parallel}=1$ the energy curve is
almost flat, although it does not show a minimum. 
A minimum at $n_{\parallel} \neq 1$ will appear if repulsive short-range 
interactions are taken into account. The energy gain we
obtained for the representative value of $J/t$ is about $0.25t\simeq
0.6J$ per hole. The picture shown in Fig.\ref{fig4} varies
slowly as a function of $J/t$, with the boundary of stripe stability moving
towards smaller $n_{\parallel}$ at smaller $J/t$.
Fig.\ref{fig4} (inset) shows the spatial density of holes $N(x)$ and modulus 
of the staggered magnetization $|M(x)|=|\langle S^z(x)\rangle|$ 
in the direction perpendicular
to the stripe at the linear hole density $n_{\parallel}=2/3$. To extract
these we calculated an explicit distribution of the hole and spin flips 
within the wave function of the single elementary excitation
and then integrated these quantities over $k_y$. 
Our results for $N(x)$ are in remarkable agreement with the same
quantity for the single stripe obtained numerically in Ref.\cite{WS} for
the case of the isotropic $t$~-~$J$ model. 

Another issue to address is the stability of an ADW to
the untwisting of the order parameter. In other words, it is necessary to
show that the $\pi$-shift of the order parameter is stable against 
small distortions. 
Energy change of the transverse hole fluctuations by the untwist is 
$\Delta E_{tr}\sim +A(n_{\parallel})\varphi^2$, $\varphi$
being the angle of the untwist. On the other hand, the energy of magnetic
frustration caused by an ADW is decreased by the untwist, 
$\Delta E_w=-B(n_{\parallel})\varphi^2$ with the actual numbers showing 
that $A>B$ only at large hole densities $n_{\parallel}\sim 1$. 
At the same time, the impact of the small distortion on the
longitudinal hole movement is more dramatic. When the phase shift
across the wall is different from
$\pi$ the hole motion along the wall will lead to the tail of weak
spin defects with energy
$J\varphi^2/4\ll J$ per unit length and the hole will be
localized on the lengthscale
$l_{loc}\sim (J\varphi^2/t)^{-2/3}\gg 1$. The corresponding
energy increase due to the untwist is 
$\Delta E_{l}\sim +C\varphi^{4/3}$, which {\it always}
dominates the $\varphi^2$-term at small $\varphi$ and ensures the
{\it local} stability of the $\pi$-shift at {\it all} hole densities.

In the case of a hole in a pure Ising background it is known that
the hole can escape from the confinement potential via some high energy
processes \cite{Trug} remaining beyond the retraceable path
approximation (Trugman loops). 
However, in the case of the hole at the ADW 
the holon in the stripe has no spin, whereas the
hole in the bulk has both spin and charge. Therefore, to leave the
stripe and acquire a spin the hole must create a spinon
which costs energy $\sim J$. In other words, a holon can 
only virtually decay into a spin-polaron and a spinon.

In the more realistic $t$~-~$J$ model spins are dynamic, which would
renormalize the energies of the magnetic background and the holes in
our picture. We believe, however, that the essential physics of the
system will remain the same.
 
In conclusion, we have presented an analytical study of a stripe of
holes at an ADW of AF spins. Longitudinal as well as transverse
kinetic energy of holes are explicitly taken into account in our
approach, and their role in stabilization of the stripe
as a ground state of the system is revealed. We have provided a description
of the charge carriers building the stripe as a system of 1D
elementary excitations, unifying the features of holons, spinons, and AF
spin polarons. This represents a new level of 
 understanding of the structure of the stripe phase in cuprates.

We would like to acknowledge fruitful discussions with A.~Balatsky,
C.~Di~Castro, G.~Ortis, D.~Poilblanc, S.~Trugman, S.~White, and J.~Zaanen.
We thank the partial support provided by 
a CULAR research grant under the auspices of
the US Department of Energy. A.~H.~C.~N. acknowledges support from the
Alfred P.~Sloan foundation.


\onecolumn
\noindent
\noindent
\end{document}